\documentclass[12pt]{article}
\usepackage{amstex}
\usepackage{epsfig}
\title{
\bf The Phase Diagram of the Gonihedric\\ 
3d Ising Model via CVM}
\author{ {\it E.N.M. Cirillo, G. Gonnella} \\
         Dipartimento di Fisica
         dell'Universit\`a degli Studi di Bari and\\
	 Istituto Nazionale di Fisica Nucleare, Sezione di Bari\\
         via G. Amendola 173, I-70126 Bari, Italy
        \\
        \\
        {\it D.A. Johnston} \\
         Dept. of Mathematics Heriot-Watt University\\
	 Riccarton Edinburgh, EH14 4AS, Scotland
	\\
	\\
	{\it A. Pelizzola} \\
         Dipartimento di Fisica del Politecnico di Torino and\\
	 Istituto Nazionale di Fisica per la Materia,\\
         c. Duca degli Abruzzi 24, 10129 Torino, Italy}
\begin{document}
\maketitle
                      {\Large
                      \begin{abstract}
%
We use the
cluster variation method (CVM)
to investigate the phase structure 
of the 3d gonihedric
Ising actions defined by Savvidy and Wegner.
The geometrical spin cluster boundaries
in these systems serve as models
for the string worldsheets
of the gonihedric string embedded in ${\bf Z}^3$.  
The models are interesting from
the statistical mechanical point
of view because they have a vanishing bare surface tension. 
As a result 
the action depends only
on the angles of the discrete surface and not 
on the area, which is the antithesis of the
standard 3d Ising model.

The results obtained with the CVM are in good agreement
with Monte Carlo simulations for the critical temperatures
and the order of the transition as the self-avoidance
coupling $\kappa$ is varied. The value of the magnetization
critical exponent $\beta = 0.062 \pm 0.003$, calculated with the
cluster variation--Pad\`e 
approximant method, is also close to the simulation results.
%
                        \end{abstract} }
%
  \thispagestyle{empty}
%
%
  \newpage
%
                  \pagenumbering{arabic}

\section{The Model}
In this paper we discuss the use of the cluster variation method
(CVM) in mapping out the phase diagram of the
gonihedric 3d Ising model. 
The gonihedric 3d Ising model is a generalization 
of the usual 3d Ising model where planar Peierls 
boundaries between $+$ and $-$ spins can be created
at zero energy cost. It has been introduced in \cite{1}
in relations with string theory.
The CVM method in other contexts has shown itself
to be an  accurate and economical way of describing the phase
diagram  both for first order and continuous transitions.
As both the model and the 
method may be unfamiliar, we outline both in turn before
going on to describe our results.

The genesis of the model is in
a novel discretized
random surface theory, the so-called gonihedric string \cite{1},
\begin{equation}
S = {1 \over 2} \sum_{\langle ij\rangle } | \vec X_i - \vec X_j | \theta (\alpha_{ij}),
\label{e4a}
\end{equation}
where the sum is over the edges of some triangulated surface,
$\theta(\alpha_{ij}) = | \pi - \alpha_{ij} |^{\zeta}$,
$\zeta$ is some exponent,
and $\alpha_{ij}$ is the dihedral angle between
neighbouring triangles with common link $\langle ij\rangle $.
This definition of the action was
inspired by the geometrical notion of
the linear size
of a surface, as originally defined by Steiner \cite{steiner}.

In eq.\ (\ref{e4a}) the surface itself is
discretized, rather than the space in which it is embedded.
An alternative approach
to discretizing the linear size would be to
restrict the allowed surfaces to the plaquettes of a (hyper)cubic lattice,
which corresponds to also discretizing the target space.
Savvidy and Wegner \cite{7,8,8a,8b} did this and
rewrote the resulting model
as an equivalent generalized Ising model using
the geometrical spin cluster boundaries
to define the surfaces.
The energy of a surface on a cubic
lattice is then given by
$E=n_2 + 4 \kappa n_4$, where $n_2$ is the number
of links where two plaquettes meet at a right angle,
$n_4$ is the number of links where four plaquettes
meet at right angles, and $\kappa$ is a free
parameter which determines the relative
weight of a self-intersection of the surface.
In the limit $\kappa \rightarrow \infty$
the surfaces are strongly self-avoiding,
whereas the opposite limit $\kappa \rightarrow 0$
is that of ``phantom'' surfaces that 
pass through themselves without any energy penalty.
It is worth emphasizing
that the energy is very different from that of
the standard 3d Ising
model with nearest neighbour
interactions where the surfaces are weighted 
entirely by their areas and not at all by their embeddings.
 
On a cubic lattice
the generalized gonihedric Ising hamiltonian which
reproduces the energy
$E=n_2 + 4 \kappa n_4$ contains nearest neighbour ($\langle i,j\rangle $),
next to nearest neighbour ($\langle \langle i,j\rangle \rangle $) and round a plaquette ($[i,j,k,l]$)
terms
\begin{equation}
{\bf -}H = 2 \kappa \sum_{\langle ij\rangle }\sigma_{i} \sigma_{j} -
\frac{\kappa}{2}\sum_{\langle \langle i,j\rangle \rangle }\sigma_{i} \sigma_{j} 
+ \frac{1-\kappa}{2}\sum_{[i,j,k,l]}\sigma_{i} \sigma_{j}\sigma_{k} \sigma_{l}.
\label{e1}
\end{equation}
Such generalized Ising actions and their
equivalent surface formulations have quite complicated phase structures
for generic choices of the couplings \cite{9,9a,10}. The particular
ratio of couplings in eq.\ (\ref{e1}), however, is special
and introduces a novel
symmetry into the model, related
to a zero-temperature high degeneracy point where
it is possible to flip any plane
of spins at zero energy cost.

Various approaches have been used to investigate these
models, including a zero-temperature analysis,
mean-field theory and Monte Carlo simulations.
We briefly outline the results thus obtained
for comparison with our CVM calculation in this paper.

In an analysis of the zero temperature
ground states of the model
we write the full lattice
hamiltonian as a sum over individual cube hamiltonians
$h_c$
and observe that if the lattice can be tiled by
a cube configuration minimizing the individual $h_c$
then the ground state energy density is
$\epsilon_0 = {\rm min}\;  h_c$ \cite{9}.
This approach confirms that a layered ground state
with parallel layers of flipped spins 
perpendicular to one of the lattice axes and
arbitrary interlayer spacing is degenerate with
the ferromagnetic ground state for all $\kappa$.
In addition, at $\kappa=0$ extra
ground states appear  where all the spins of one
of the 2 sublattices of the cubic lattice are flipped
\cite{9,11}.
 
In the mean field approximation the spins
are replaced by average site magnetizations.
The calculation
of the mean field free energy can be performed
by considering independent magnetizations for the 8 sites
of an elementary cube. Therefore, 
as in the
zero temperature approach,
the energy can be  still decomposed into a sum of individual cube terms.
Numerical iteration of the resulting 
eight coupled mean-field equations 
shows a single transition from
a paramagnetic high temperature state to a layered, or 
the equivalent ferromagnetic, low temperature state \cite{11}.
The inverse critical temperature $\beta_c$ determined in this fashion decreases
quite sharply with $\kappa$.

The flip symmetry of the model 
poses something of a problem when carrying out
Monte Carlo simulations. 
A
simple ferromagnetic order parameter such
as the magnetization
\begin{equation}
M = \left\langle  {1 \over L^3} \sum_i \sigma_i \right\rangle 
\label{ord}
\end{equation}
will be zero in general, because of the layered
nature of the ground state. Staggered
magnetizations also fail as order parameters because the interlayer
spacing can be arbitrary.
In \cite{11} 
boundary conditions have been suitably chosen in order
to pick out the ferromagnetic 
ground state allowing the use of standard
(unstaggered) magnetization to extract
magnetic critical exponents.
Monte Carlo simulations with such boundary conditions \cite{11} for different
$\kappa$ values on lattices of various sizes
allowed a finite size scaling analysis to be carried
out in order to extract estimates for some of the critical
exponents.
For $\kappa=1$ this gave $\nu = 1.2(1)$
from the ratios of slopes of Binder's magnetization cumulants,
$\gamma / \nu = 1.79(4)$ from the FSS of the susceptibility
$\chi$,
and $( \alpha - 1 ) / \nu = -1.3(2)$
from the FSS of the energy with 
two different sorts of fixed boundary conditions.
All these exponents, rather remarkably given that the
model is defined in three dimensions, 
are close to the Onsager values of the two-dimensional
Ising model with nearest neighbour interactions,
as was the critical temperature $\beta_c=0.44$.

Simulations of other
$\kappa \ge 1$ values gave very similar
results. However, the $\kappa=0$ model \cite{12}
\begin{equation}
H= \frac{1}{2}\sum_{[i,j,k,l]}^{ }\sigma_{i} \sigma_{j}\sigma_{k} \sigma_{l}.
\label{e2}
\end{equation}
appeared to be a special case, displaying a first order
transition. The transition stayed
first order at $\kappa=0.1$ but softened rapidly
as $\kappa$ increased, so the crossover to the second
order behaviour seen at $\kappa=1$ was quite sharp.

\section{The Cluster Variation Method}

The cluster variation method, or CVM for short, is based
on a truncation of the cluster (cumulant) expansion
of the free energy density functional on which the variational
formulation of statistical mechanics is based \cite{kikuchi,CVM2}. 
Unlike mean field theory it generally locates rather accurately the
boundaries between different phases in complex phase diagrams and, 
using the recently proposed cluster variation--Pad\`e approximant
method \cite{cvpam1,cvpam2,cvpam3} one can extract 
non-classical, precise estimates of the critical exponents.

For a generic Ising-like model described by a hamiltonian $H$ 
on a lattice $\Lambda$ the exact free energy can in principle be obtained 
by minimizing the functional 
\begin{equation}
F[\rho_\Lambda] = {\rm Tr}  \left(\rho_\Lambda \; H  
+ \frac{1}{\beta} \rho_\Lambda \ln \rho_\Lambda \right), 
\end{equation}
of the trial density matrix $\rho_\Lambda$, 
subject to ${\rm Tr} ( \rho_\Lambda ) = 1$. 

Assuming that the hamiltonian can be written as a sum of cluster
contributions 
\begin{equation}
H = \sum_{\alpha \in \Gamma} h_{\alpha},
\end{equation}
where $\Gamma$ is a collection of clusters
that suffices to enumerate
the interactions, an approximate free energy functional can be
written in the form \cite{CVM2}
\begin{equation} 
F = \sum_{\alpha \in \Gamma} {\rm Tr}  \left(\rho_{\alpha} \; h_{\alpha} \right) + \frac{1}{\beta} 
\sum_{\alpha \in P} a_{\alpha} {\rm Tr} ( \rho_{\alpha} \ln \rho_{\alpha} )
\label{trunc}
\end{equation}
where $P$ is a suitable set of clusters (which must contain $\Gamma$
as a subset), the largest of which (called maximal clusters) 
have to reflect in some way the
symmetry of the lattice, 
and the $a_{\alpha}$ are numerical coefficients determined by 
\begin{equation}
\sum_{\alpha \subseteq \beta \in P} a_{\beta} =1, \qquad \forall
\alpha \in P.
\end{equation}
The constraints are now
\begin{eqnarray}
Tr ( \rho_{\alpha} ) &=& 1  , \; \; \; \; \; \; \; \; \; \; \; \;
 \; \; \; \; \; \alpha \; \in \; P \nonumber \\
   \rho_{\alpha} &=& {\rm Tr}_{\beta \setminus \alpha} ( \rho_{\beta} ) , 
\; \; \; \alpha \subseteq \beta
\end{eqnarray} 
and the latter can actually be used as a definition of the density
matrices of the subclusters of the maximal clusters.

An important feature of the CVM is that the local minima of the
approximate free energy can be easily found by means of a simple
iterative procedure called natural iteration method \cite{CVM1,CVM3},
which has the property that, for any given initial set of density
matrices, the iteration always converges to a local minimum of the
free energy. 

In the present work, we have used the cube approximation of the
CVM, that is the maximal clusters of the set $P$ are the elementary
cubic cells of our simple cubic lattice; for this approximation the
free energy density functional has the form (see \cite{kikuchi} for an
application to the simple nearest neighbour Ising model)
\begin{eqnarray}
f[\rho_8] &=& {\rm Tr} (\rho_8 H_8) + \frac{1}{\beta} \Bigg[
{\rm Tr} {\cal L} (\rho_8) 
- \frac{1}{2} \sum_{\rm plaquettes} 
{\rm Tr} {\cal L} (\rho_{4,{\rm plaquette}}) \nonumber \\
&& + \frac{1}{4} \sum_{\rm edges}
{\rm Tr} {\cal L} (\rho_{2,{\rm edge}}) - \frac{1}{8} \sum_{\rm sites}
{\rm Tr} {\cal L} (\rho_{1,{\rm site}}) \Bigg],
\label{fcube}
\end{eqnarray}
where $H_8$ is the contribution of a single cube to the hamiltonian
(when splitting the total hamiltonian $H$ into single cube
contributions one has to keep in mind that nearest neighbour
interactions are shared by four cubes 
and then will get a coefficient 1/4 in $H_8$, and similarly
next-nearest neighbour and plaquette interactions will get a
coefficient 1/2), ${\cal L}(x) = x \ln x$, $\rho_\alpha$ with $\alpha
= 8$ (4, 2, 1) denotes the cube (respectively plaquette, edge, site)
density matrix, and the sums in the entropy part are over all
plaquettes (edges, sites) of a single cube. Notice that we have not
assumed any {\it a priori} symmetry property for our density matrices.

The cluster variation method can be viewed as a generalized mean field
theory, and hence it is clear that it can give only classical
predictions for the critical exponents. In order to overcome this
difficulty, one can use the recently proposed cluster
variation--Pad\`e approximant method (CVPAM)
\cite{cvpam1,cvpam2,cvpam3}, which has proven to be a rather accurate
technique, although not very demanding in terms of computer time. The
basic idea of the CVPAM is that, since the CVM gives, for Ising-like
models, very accurate results at low and high temperatures (i.e.\ far
enough from the critical point), one can try to extrapolate this
results in order to study the critical behaviour. In order to
determine the critical exponent of the order parameter $m$, for example,
one calculates $m(\beta)$ with the CVM up to a temperature at
which the error can be estimated to be very small (typically of order
$10^{-5}$), and then constructs, by a simple interpolation, Pad\`e
approximants for the logarithmic derivative of $m(\beta)$: the pole
and the corresponding residue of each Pad\`e approximant are then
estimates for the critical temperature and for the critical exponent
respectively. 

\section{The Results}
In this section we describe our results for the phase diagram of the 
model (\ref{e1}) at different values of the parameter $\kappa$.
When the temperature is lowered, the model (\ref{e1}) undergoes a phase 
transition towards a low temperature ordered phase: this transition has been 
investigated by means of the CVM in the cube approximation described in the 
preceding section.
\par
\begin{figure}[t]
\begin{center}
\mbox{\epsfig{file=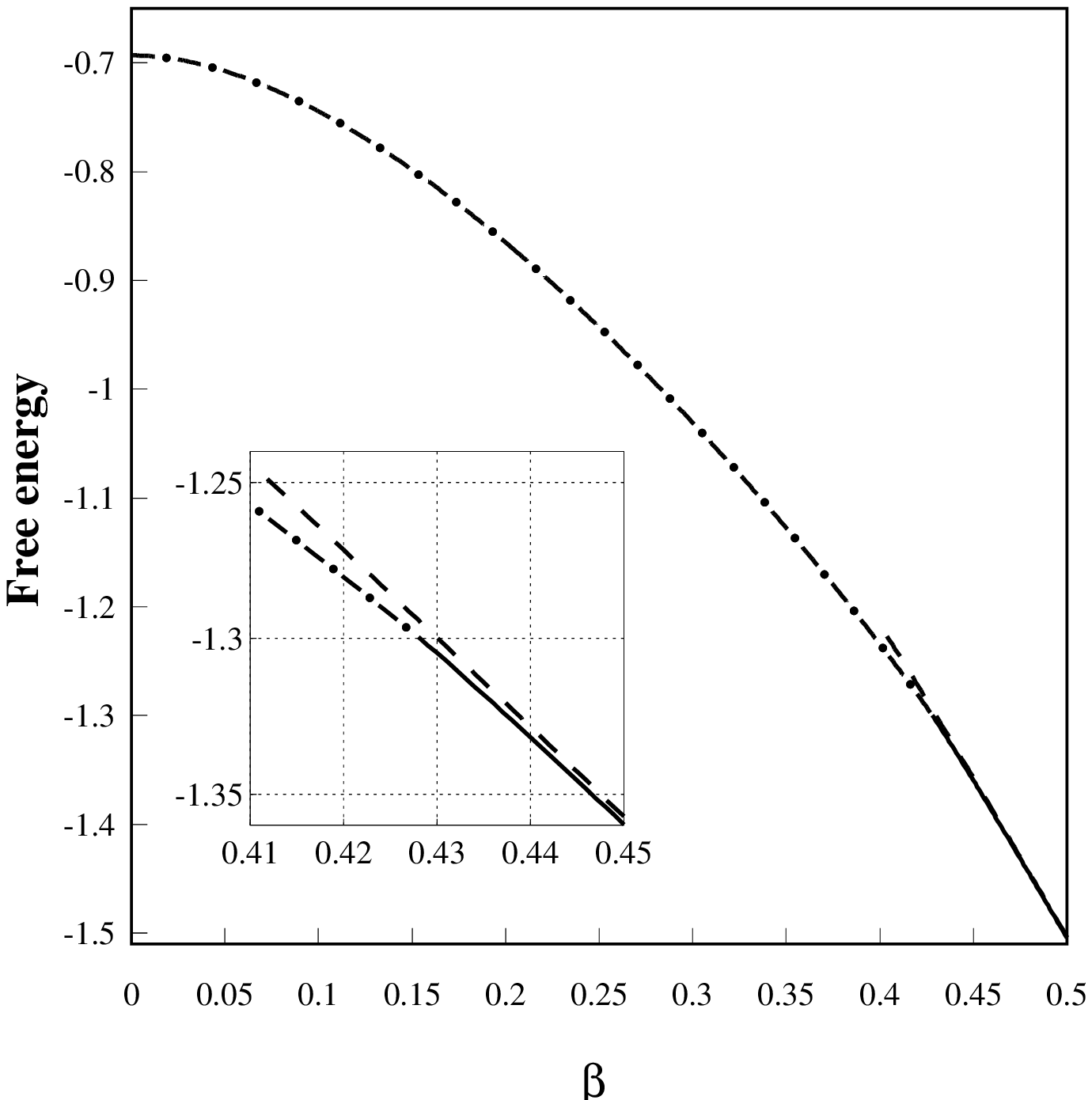,width=8cm,height=8cm}}
\end{center}
\caption{Solid, dash-dotted and dashed lines represent respectively the free
energies $f_F(\beta)$, $f_P(\beta)$ and $f_L(\beta)$ at $\kappa=1$.}
\label{free-ener}
\end{figure}

First of all we describe our results at $\kappa=1$. We have solved
numerically our approximate variational principle
for $\beta$ ranging in the interval
$0\le\beta\le 0.5$ and we have calculated the values $f_P(\beta)$,
$f_F(\beta)$ 
and $f_L(\beta)$ of the local minima of the free energy corresponding 
respectively to the paramagnetic, ferromagnetic and layered phases. The three 
functions $f_P(\beta)$, $f_F(\beta)$ and $f_L(\beta)$ are plotted in Fig. 
\ref{free-ener}.
\par
When $\beta < 0.404$ the unique local minimum is that 
corresponding to the paramagnetic phase; in the interval 
$0.404\le\beta <0.427$ we find both the paramagnetic and layered minima of 
the free energy and $f_P(\beta)<f_L(\beta)$ in the whole interval. At low 
temperature when $\beta\ge 0.427$ the ferromagnetic local minimum appears and 
it happens to be the global minimum of the free energy. Therefore the CVM 
predicts that the zero temperature degeneracy between phases with all possible 
sequences of ``$+$" and ``$-$" planes is broken at finite temperature.
\par
The transition at $\beta_c=0.427$ from the paramagnetic to the
low temperature ferromagnetic phase is critical.
Indeed, when the inverse temperature $\beta$ is
lowered below $\beta_c$, the ferromagnetic local minimum of the free energy 
disappears. 
\par
\begin{figure}[b]
\begin{center}
\mbox{\epsfig{file=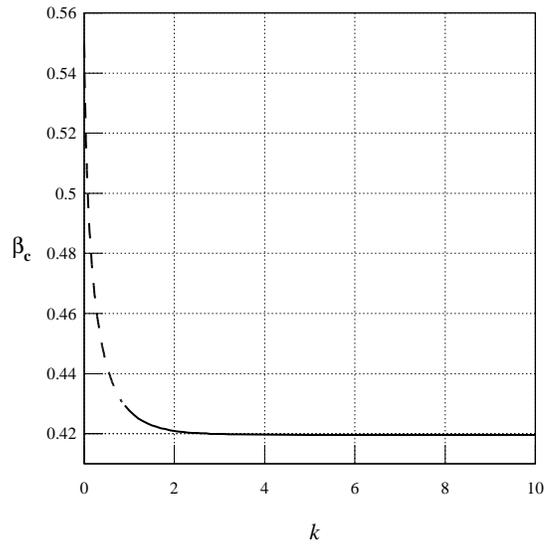,width=8cm,height=8cm}}
\end{center}
\caption{The inverse transition temperature is plotted as a function of the
parameter $\kappa$. Dashed and solid lines represent respectively first and
second order inverse transition temperatures.}
\label{kbeta}
\end{figure}

We have studied the model (\ref{e1}) for other values of the parameter
$\kappa$ 
ranging in the interval $0\le \kappa \le 10$. For each $\kappa>0$ we have
found that at  
low temperature the model is in the ferromagnetic phase and we have calculated 
the inverse transition temperature $\beta_c$; in Fig. \ref{kbeta} we have 
plotted $\beta_c$ as a function of $\kappa$. 
\par
At sufficiently low values of $\kappa$, that is
$\kappa<\kappa_{\rm tr}=0.87\pm 0.01$, the nature of the 
transition changes over to a first order behaviour which is strengthened
as $\kappa$ is lowered.
\par
In Table \ref{tab} our results for $\beta_c$ 
are compared with Monte Carlo and mean field approximation results obtained in 
\cite{11}; Monte Carlo and CVM predictions are in good agreement.
\begin{table}[h]
\begin{center}
{\bf Table 1}
\end{center}
\vskip 1 truecm
\begin{center}
\begin{tabular}{lccccccc} \hline\hline
 & & & & & & & \\
 & & & &$\kappa$& & & \\ 
 & & & & & & & \\
 &0&0.25&0.5&1&2&5&10 \\ \hline
 & & & & & & & \\
$\beta_c$ CVM&0.550&0.464&0.443&0.427&0.421&0.420&0.420\\ 
 & & & & & & & \\
$\beta_c$ MC$^{\cite{11,12}}$&0.505&-&0.44&0.44&0.44&0.44&0.44\\ 
 & & & & & & &\\ 
$\beta_c$ MF$^{\cite{11}}$&0.325&0.31&0.278&0.167&0.09&0.0335&0.02\\
 & & & & & & &\\ \hline\hline
\end{tabular}
\vskip 1.5 truecm
\caption{
Our results for the inverse transition temperature $\beta_c$ at 
different values of the parameter $\kappa$ are listed. These CVM results are 
compared with Mean Field and Monte Carlo previous results.}
\label{tab}
\end{center}
\end{table}
\par
\begin{figure}[bhp]
\begin{center}
\mbox{\epsfig{file=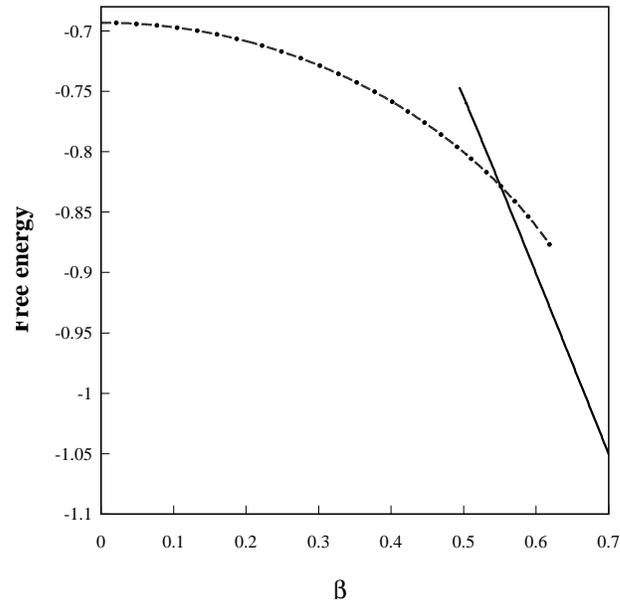,width=9cm,height=9cm}}
\end{center}
\caption{The dash-dotted and the solid lines represent respectively the free 
energy of the paramagnetic phase and of the coexisting low temperature phases 
at $\kappa=0$.}
\label{kzero}
\end{figure}

At $\kappa=0$ the first order transition is at $\beta_c=0.550$.
Moreover at low  
temperature the layered phases, the ferromagnetic phase and their 
antiferromagnetic versions obtained by flipping the site-magnetizations of one 
of the 2 sublattices of the cubic lattice coexist with the same free energy.
In Fig. \ref{kzero} the free energy of 
the paramagnetic phase (dash-dotted line) and of the 
coexisting ordered phases (solid line) are depicted.
\par
Therefore, at $\kappa=0$ the zero temperature symmetry of the model is not 
violated, and the CVM correctly respects the ferro-antiferromagnetic exact 
symmetry of the partition function.
\par
Finally, for some values of $\kappa$ such that the transition is
critical, that is 
$\kappa=1,2,5,10$, we have evaluated the magnetization critical
exponent $\beta$ by means of the cluster variation-Pad\`e approximant
method \cite{cvpam1,cvpam2,cvpam3}, obtaining $0.059 \le \beta \le
0.062$ at $\kappa = 1$ and $\beta \simeq 0.065$ at $\kappa = 2, 5,
10$. These results clearly suggest that the exponent is independent of
$\kappa$ and a prudential estimate is $\beta = 0.062 \pm 0.003$, which
has to be compared with the conjectured Onsager value 1/8 on one side,
but also 
with the estimate, based on finite size scaling of Monte Carlo
results \cite{11}, $\beta/\nu = 0.04(1)$: given the Monte Carlo
estimate $\nu = 1.2(1)$ one can say that the $\beta$ values predicted
by CVM and simulations are in rather remarkable agreement. 

\section{Conclusions}
In this work we have applied the cube approximation
of the Cluster Variation Method  to find
the phase diagram of the gonihedric $3d$ Ising model
defined in eq. (\ref{e1}).
Moreover the low-temperature CVM results for the order parameter
have been used to evaluate the exponent $\beta$ at $\kappa=1,2,5,10$
via the cluster variation--Pad\`e approximant method.

We summarize here our main results.
The CVM approximation gives 
values of the inverse critical temperature 
in quite good agreement with those predicted 
by Monte Carlo simulations. 
The transition remains critical 
for $ \kappa>\kappa_{\rm tr}=0.87\pm 0.01$, where it becomes of first order.
Our evaluation of the critical exponent 
$\beta = 0.062 \pm 0.003$ of the magnetization is also close to the 
simulations results of \cite{11}.

The new result of this paper is that the CVM 
predicts at finite temperatures a violation of the symmetry 
of the hamiltonian (\ref{e1}). We find that in the ordered
region of the model  the ferromagnetic phase is always  stable 
with respect to the lamellar phase. This suggests
to study the model (\ref{e1}) in a parameter space larger
than the one used in this paper. The knowledge of the global topology
of the phase diagram in an enlarged parameter space \cite{prep}
could be useful to answer the main question 
set by the Monte Carlo results and confirmed by the results
of this paper about the nature of the critical transition
of the model (\ref{e1}) at sufficiently high values of $\kappa$.


\clearpage\newpage
%
%

\clearpage\newpage

\end{document}